\documentclass[fleqn,twoside]{article}
\usepackage{espcrc2}
\usepackage{graphicx}

\usepackage{graphicx}
\usepackage[figuresright]{rotating}

\newcommand{\AmS}{{\protect\the\textfont2
  A\kern-.1667em\lower.5ex\hbox{M}\kern-.125emS}}

\newcommand{\degrees}{\mbox{${^\circ}$}}
\newcommand{\fbinv}{\mbox{$\rm fb^{-1}$}}
\newcommand{\phis}{\mbox{$\phi_s$}}
\newcommand{\sphis}{\mbox{$\sigma({\phi_s})$}}
\newcommand{\phisphiphi}{\mbox{$\phi_s(\phi\phi)$}}
\newcommand{\sphisphiphi}{\mbox{$\sigma(\phi_s(\phi\phi))$}}
\newcommand{\sgamma}{\mbox{$\sigma({\gamma})$}}

\newcommand{\Bs}{\mbox{$B^0_s$}}
\newcommand{\Bstojpsiphi}{\mbox{$B^0_s \to J/\psi\phi$}} 
\newcommand{\Bstophiphi}{\mbox{$B^0_s \to \phi\phi$}} 
\newcommand{\Bdtojpsiks}{\mbox{$B^0 \to J/\psi K^0_S$}} 
\newcommand{\Bdtophiks}{\mbox{$B^0 \to \phi K^0_S$}} 

\newcommand{\Btodk}{\mbox{$B \to D K$}} 
\newcommand{\Bstodsk}{\mbox{$B^0_s \to D_s^{\mp} K^{\pm}$}}
\newcommand{\Bdtokstarmumu}{\mbox{$B^0 \to K^{*0} \mu^+ \mu^-$}}

\title{LHCb upgrade plans}
\author{F. Muheim%
        \address{School of Physics, University of Edinburgh,
        Mayfield Road, Edinburgh EH9 3JZ, United Kingdom}
  \\On behalf of the LHCb Collaboration
       }

\begin{document}

\begin{abstract}
The LHCb experiment will operate for about five years 
at a luminosity of $2 \times 10^{32}\,\rm cm^{-2} s^{-1}$ 
and plans are to accumulate a data sample of $\sim 10\;\rm fb^{-1}.$
Here we present the physics programme and detector design for a future high
luminosity phase of the LHCb experiment. An upgraded LHCb experiment would operate
at ten times the design luminosity, i.e. at  $\sim 2 \times 10^{33}\,\rm cm^{-2} s^{-1}$ 
and aims to collect a data sample of $\sim 100\; \rm fb^{-1}$ over five years.
This programme would allow the probe of new physics at an unprecedented level.
Key measurements include the $B^0_s$ mixing phase $\phi_s$ in 
\Bstojpsiphi\ and \Bstophiphi\ decays
with a significant sensitivity to the small Standard Model prediction and   
a very precise measurement of the CKM angle~$\gamma$ in tree diagram decays. 
Initial studies of the modified LHCb trigger and detectors are 
presented. 
The upgraded LHCb experiment can run with or without an LHC luminosity upgrade.
\vspace{1pc}
\end{abstract}

\maketitle

\section{INTRODUCTION}
\label{sec:intro}
   
The Standard Model (SM) is a very successful effective theory. 
Over the last few years it has emerged that 
the CKM mechanism is likely the dominant mechanism and new physics (NP)   
beyond the SM would appear as corrections.
If NP were observed at the LHC its flavour sector must be studied.
Here we discuss the plans for an upgraded LHCb experiment.
In Section~\ref{sec:lhcb} we present the LHCb experiment and physics programme, 
and motivate running at higher luminosities.
The physics reach of a 100~\fbinv\ data sample 
is discussed in Section~\ref{sec:physics} for selected key measurements: 
the $B^0_s$ mixing phase \phis\ from 
the tree decay $B^0_s \to J/\psi\phi$, the $b\to s$ penguin decay $B^0_s \to \phi\phi$, as well
as the CKM angle~$\gamma$.
In Section ~\ref{sec:detector} we present the plans for upgrading the LHCb trigger and
detectors. The report ends with conclusions.

\section{THE LHCb PHYSICS PROGRAMME}
\label{sec:lhcb}

\subsection{The LHCb experiment}

The goal of the LHCb experiment is to perform precise measurements of CP violation 
and rare decays of $B$ mesons. 
The LHCb detector is a forward arm spectrometer which is equipped with 
state-of-the-art vertexing capabilities, and 
is designed to trigger efficiently on muons, electrons and photons 
as well as hadronic final states.
Two Ring Imaging Cherenkov (RICH) detectors allow  excellent 
charged particle identification. 
The trigger and flavour tagging of LHCb are discussed  
elsewhere in these proceedings~\cite{ref:rodrigues,ref:ruiz}.
The LHCb experiment will start taking data in 2007 
at the Large Hadron Collider (LHC) at CERN.
More details about the current status of LHCb and its first year data-taking
plans are described in~\cite{ref:garrido,ref:corti}.

\subsection{LHCb - the first five years}

The LHCb experiment is designed to operate at a luminosity 
${\cal L} \sim 2 \times 10^{32}\, \rm  cm^{-2}\,s^{-1}$. This choice maximises
the number of bunch crossings with exactly one $p p$ interaction and
reduces the radiation damage to the vertex detector. 
This luminosity is a factor of $\sim 50$ below the LHC design value 
which is achievable by using a tunable amplitude function, $\beta^*$, 
at the LHCb interaction point. 
Therefore this luminosity will very likely be
reached during the first LHC physics run.
In a nominal year of $10^{7}\,\rm s$, a data sample of $2\;\rm fb^{-1}$ 
will be collected by LHCb.

\begin{table*}[htb]
\caption{Expected signal yields $S$, signal to background ratios $B/S$ and sensitivities 
for $2\; \rm fb^{-1}$ of data. 
The yields also include charge conjugate modes, which are implied throughout.
The parameters are defined in the text; $C_7^{\rm eff}/C_9^{\rm eff}$ is the ratio of Wilson coefficients
and $A_{CP}$ is the asymmetry in direct CP violation.
}
\label{tab:sensitivities}

\renewcommand{\tabcolsep}{1pc} 
\renewcommand{\arraystretch}{1.1} 

\begin{tabular}{@{}clccc}
\hline
  &  Decay                                        &     Yield $S$ &  $B/S$    & Precision     \\
\hline
$\gamma$ & $B^0_s \to D_s^{\mp}K^{\pm}$   &     5.4k   &  $< 1.0$  & $\sigma(\gamma) \sim 14\degrees$ \\
 & $B^0 \to \pi^+\pi^-$                            &     36k    &  0.46     & $\sigma(\gamma) \sim 4\degrees$ \\
 & $B^0_s \to K^+ K^- $                              &     36k    &  $<0.06$  &       \\
 & $B^0 \to D^0(K^-\pi^+,K^+\pi^-)K^{*0}$          & 3.4k, 0.5k &  $<0.3, < 1.7$ &$\sigma(\gamma) \sim 7\degrees - 10\degrees$ \\ 
 & $B^0 \to D^0(K^+K^-,\pi^+\pi^-)K^{*0}$          &  0.5k      &  $<1.4$   &  \\
 & $B^- \to D^0(K^- \pi^+, K^+\pi^-) K^-$            &  56k, 710  & $0.6, 1.5 - 4.3$     & $\sigma(\gamma) \sim 5\degrees - 15\degrees$ \\
 & $B^- \to D^0(K^+K^-/\pi^+\pi^-) K^-$              &     7.6k   &  1.0      &  \\
 & $B^- \to D^0(K^0_S \pi^+\pi^-) K^-$               &   1.5 - 5k &  $< 0.7 - 2.3$      & $\sigma(\gamma) \sim 8\degrees - 16\degrees$ \\ 
\hline
$\alpha$  & $B^0 \to \pi^+\pi^-\pi^0$              & 14k        & $<0.8$    & $\sigma(\alpha) \sim 10\degrees$ \\
 & $B^{+,0} \to \rho^+\rho^0, \rho^+\rho^-, \rho^0\rho^0$  & 9k, 2k, 1k & $1, <5, < 4$  & \\
\hline
$\beta$ & \Bdtojpsiks\                               & 216k       & 0.8       & $\sigma(\sin 2\beta) \sim 0.022$ \\
\hline
$\Delta m_s$ & $B^0_s \to D_s^- \pi^+$               & 80k        & 0.3       & $\sigma(\Delta m_s) \sim 0.01\; \rm ps^{-1}$ \\
$\phi_s$     & \Bstojpsiphi\                         & 131k       & 0.12      & $\sphis \sim 0.023$ rad \\       
\hline
Rare~~  &   $B^0_s \to \mu^+ \mu^-$                  & 17         & $<5.7$    & \\ 
Decays  &   \Bdtokstarmumu                           & 4.4k       & $< 2.6$   & $\sigma(C_7^{\rm eff}/C_9^{\rm eff}) \sim 0.13$\\
 &   $B^0 \to K^{*0} \gamma$                       & 35k        & $<0.7$    & $\sigma(A_{CP}) \sim 0.01$ \\
 &   $B^0_s \to \phi \gamma$                         & 9.3k       & $<2.4$    & \\
\hline
\end{tabular}\\[2pt]
\end{table*}

The large cross section of $500\,\mu$b for the production 
of $b \bar b$-quark pairs at the LHC will allow LHCb to collect much larger
data samples of selected $B$ meson decays than previously available.
The sensitivities of these data-sets to 
CP violating asymmetries and to other physics observables
have been studied with simulated events.
In Table~\ref{tab:sensitivities} we present a selection of signal yields $S$, 
background to signal ratios $B/S$ and precisions
on CKM angles $\gamma$, $\alpha$, $\beta$ and $\phi_s$  and other observables.
Detailed presentations on many of the channels are presented elsewhere in these
proceedings~\cite{ref:carbone,ref:xie,ref:robbe,ref:magini,ref:smizanska,ref:decapua,ref:muresan}.

The plan is to operate the LHCb experiment for five years at the design
luminosity and to collect a data sample of $6$ to $10\;\rm fb^{-1}$.
A major goal is to exploit the physics in the $B_s$ system. This includes
the observation of CP violation in $B_s$ mesons and precision measurements
of the mass difference between the $B^0_s$ mass eigenstates,
 $\Delta m_s$, and the lifetime difference $\Delta \Gamma_s$.
Other main aims are to improve the error on the CKM angle~$\gamma$ by a factor of five,
to probe NP in rare $B$ meson decays with electroweak, radiative and hadronic penguin modes,
and to make the first observation of the very rare decay $B^0_s \to \mu^+\mu^-$.

\subsection{LHCb at higher luminosity}
\label{sec:superlhcb}
After the first five years of operation, 
the precision of many LHCb physics results will remain limited by the 
statistical error of the collected data. 
The following questions arise:
is LHCb exploiting the full potential for $B$ physics
at hadron colliders and is there a science case for collecting even larger
data samples? 
Note that LHCb is the only dedicated heavy flavour experiment approved to run after 2010.
In the remainder of this report we will try to answer these questions.

The LHCb experiment has commenced  studying the feasibility of 
upgrading the detector such that it can operate at a luminosity  
${\cal L} \sim 2 \times 10^{33}\, \rm  cm^{-2}\,s^{-1}$, which is 
ten times larger than the design luminosity.
This upgrade would allow LHCb to collect a data
sample of about $100\; \fbinv$ during five years of running. 
This increased luminosity is achievable by decreasing 
$\beta^*$ at the LHCb interaction point. 
It does not require a LHC luminosity upgrade (Super-LHC)
as the LHC design luminosity is $10^{34}\; \rm  cm^{-2}\,s^{-1}$
although it could operate at Super-LHC. 
Thus an upgrade of LHCb could be implemented as early as 2013.
The number of interactions per beam crossing will increase to $n \sim 4$
which will require improvements to the LHCb sub-detectors and trigger.

\section{LHCb PHYSICS REACH WITH A 100~fb$^{-1}$  DATA SAMPLE}
\label{sec:physics}

\subsection{Weak mixing phase \phis} 

Flavour-changing neutral currents (FCNCs) are particularly sensitive to NP. 
New particles appear virtually in loop diagrams which lead to deviations
from SM predictions. The interference in $B^0_s$ mixing and decay to 
CP eigenstates is sensitive to the CP violating weak mixing phase $\phi_s$. 
This can be studied in the time-dependent asymmetry of flavour-tagged 
\Bstojpsiphi\ decays. The SM prediction for $\phi_s$ is very small:
$\phi_s = - 2 \chi = -2 \lambda^2 \eta \approx -0.035$\,rad
where $\lambda$ and $\eta$ are the usual Wolfenstein parameters of the CKM matrix~\cite{ref:Wolfenstein}.
Hence $\phi_s$ is a very sensitive probe of NP and will provide a 
stringent test for Non-Minimal Flavour Violation (NMFV)~\cite{ref:nmfv}.

\begin{figure}[htb]
\mbox{\includegraphics[width=0.45\textwidth]{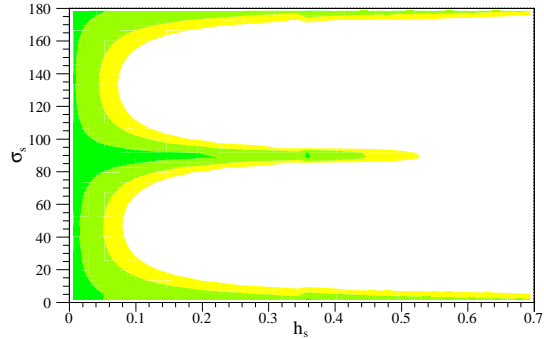}}
\caption{The allowed range for $h_s$ and $\sigma_s$ with 2 \fbinv\ of LHCb data.
The dark, medium, and light shaded areas have confidence levels $>$ 0.90, 0.32, and 0.05,
respectively, taken from~\cite{ref:ligeti}.}
\label{fig:hsss}
\end{figure}

Currently the only direct \phis\ result is from D0 who measure
$\phis = -0.79 \pm 0.56 ^{+0.14}_{-0.01}$\,rad in the untagged decay time spectrum
of \Bstojpsiphi\ decays~\cite{ref:dzero}.  
Using a full simulation,
the LHCb experiment expects to collect 131k \Bstojpsiphi\ decays
with a 2 \fbinv\ data sample~\cite{ref:fernandez}.
The expected precision on \phis\ is estimated with 
many toy Monte Carlo experiments based on the yields and resolutions obtained in the full simulation.
We obtain $\sphis \approx 0.023$\,rad where 
the \Bs\ mass difference has been set to $\Delta m_s = 20\;\rm ps^{-1}$. 
New physics in $B^0_s$ mixing can be parameterised as 
$\Delta m_s^{NP} = \left(1+ h_s e^{2i \sigma_s} \right)\Delta m_s^{SM}$
where the SM phase of $\Delta m_s^{SM}$ is equal to $-2 \chi$ 
and $h_s$ and $\sigma_s$ are, respectively, 
the amplitude and the phase of the NP contribution.
In Fig.~\ref{fig:hsss} we show the expected sensitivity for 
$h_s$ and $\sigma_s$ for LHCb with 2 \fbinv\ of data~\cite{ref:ligeti}.
With five years of data, LHCb expects to achieve a precision of $\sphis \sim 0.010$\,rad
which corresponds to about a $3\sigma$ significance for the SM value of \phis.
This precision is expected to be statistically limited, with systematic errors likely to be 
much smaller.
The precision of the SM prediction for \phis\ is better than 0.01\,rad. 

A significantly larger data-set would allow LHCb to probe new physics in $B$ mesons
at an unprecedented level of a few percent.
Here we present estimates for sensitivities with a 100~\fbinv\ data sample which
are based on scaling with luminosity. Potential trigger efficiency improvements 
are not included. While these estimates have large uncertainties, 
these are extremely useful to motivate simulation studies with the aim 
to improve or at least maintain trigger and selection efficiencies 
at ten times higher luminosities.
An upgrade of LHCb has the potential to measure the SM value of \phis\ 
with $\sim 10 \sigma$ precision ($\sphis \sim 0.003$\,rad) in \Bstojpsiphi\ decays.

\subsection{$b \to s$ transitions in \Bstophiphi }

New physics can be probed by studying FCNCs in hadronic $b \to s$ transitions.
One approach is to compare the time-dependent CP asymmetry 
in a hadronic penguin loop decay, 
where unknown massive particles could enter 
with a decay based on a tree diagram 
which generally is insensitive to NP 
and which has the same weak phase. 
The B-factories  measure  the CP asymmetry $\sin 2 \beta_{eff}$ 
in the penguin decay \Bdtophiks.  
A value for $\sin 2 \beta_{eff}$ which is different from
$\sin 2 \beta$ measured in \Bdtojpsiks\ would signal physics
beyond the SM. 
Within the current available precision, 
all $\sin 2 \beta_{eff}$ measurements are in reasonable agreement
with the SM, but all central values are lower than expected. A naive average indicates
a $2.6 \sigma$ discrepancy~\cite{ref:hfag}.

The above approach can also be applied to \Bs\ mesons and this will be exploited by LHCb.
Within the SM the weak mixing phase \phis\ is expected to be almost the same when comparing
the time-dependent CP asymmetry of the hadronic penguin decay \Bstophiphi\ with 
the tree decay \Bstojpsiphi. Due to a cancellation of the $B_s$ mixing and decay phase, 
the SM prediction for \phisphiphi\ is actually very close to zero~\cite{ref:raidal}.
Thus any measurement of $\phisphiphi \neq 0$ is a clear signal for 
NMFV.
LHCb expects to collect 1.2k \Bstophiphi\ decays in 2 \fbinv\ of data~\cite{ref:lhcbreopt}. 
By comparing with the \Bstojpsiphi\ yield we estimate a sensitivity of $\sphisphiphi \sim 0.14$\,rad 
for 10 \fbinv\ of data.
This \sphisphiphi\ precision is statistically limited. 
Scaling the sensitivity up to a data sample of 100~\fbinv\  
we estimate a precision of $\sphisphiphi \sim 0.04$\,rad. 
This sensitivity presents a very precise and exciting NP probe. 
A first-level detached vertex trigger is required to reconstruct this hadronic decay mode. 
A new LHCb simulation study on \sphisphiphi\ 
can be found in~\cite{ref:libby}.

\subsection{CKM angle $\gamma$ from  \Btodk\ and \Bstodsk}

LHCb will perform direct measurements of the CKM angle~$\gamma$ using
two interfering tree processes in neutral and charged \Btodk\ decays. 
The interference arises due to decays which are common to $D^0$ and $\bar D^0$ mesons
such as $D^0 (\bar D^0) \to K^0_S \pi^+\pi^-$ (Dalitz decay~\cite{ref:Dalitz}) and
$D^0 (\bar D^0) \to K^{\mp}\pi^{\pm}, K^+ K^-$ (ADS and GLW~\cite{ref:ADS,ref:GLW}).   
The current combined direct $\gamma$ measurements at the $B$ factories have large 
errors of about 30\degrees ~\cite{ref:hfag}.

The expected LHCb $\gamma$ sensitivities for 2~\fbinv\ of data have been estimated
for several of the above modes~\cite{ref:xie}; 
we obtain $\sgamma \sim 7\degrees - 15 \degrees$. 
In addition, LHCb will use the decay \Bstodsk\  
to measure $\gamma$ with $\sgamma \sim 14 \degrees$ in 2~\fbinv. 
By combining these measurements, LHCb will achieve a precision 
of $\sgamma \sim 5 \degrees$ in 2~\fbinv.  
The theoretical error on the SM prediction is very small.
Note that some of the $\gamma$ measurements will have a correlated 
systematic error, for example there is a Dalitz model dependence.
We take this into account when scaling to a 100~\fbinv\ data sample 
and estimate that a precision of $\sgamma \sim 1\degrees$ should be achievable.
This would allow a very precise determination of the apex of the unitarity triangle.

\subsection{Summary and comparison with the Super-B Factory}

\begin{figure}[bt]
\mbox{\includegraphics[width=0.45\textwidth]{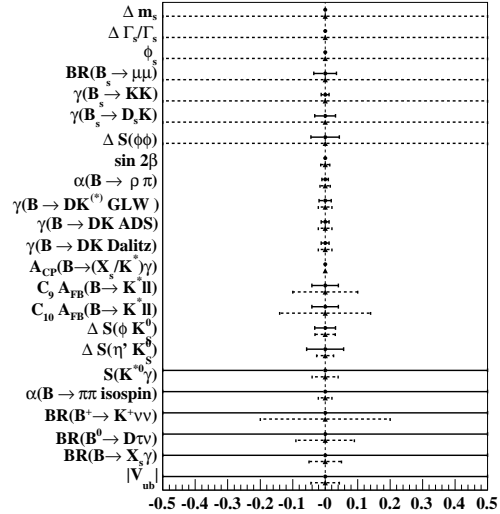}}
\caption{Expected sensitivities for LHCb (solid lines) with 
$100~{\rm fb}^{-1}$ compared to  an $e^+e^-$ Super-B factory (dashed lines) with a data sample 
of $50\;\rm ab^{-1}$~\cite{ref:superB}. 
Note that 
$\Delta S(\phi \phi) = \phisphiphi- \phis(J/\psi\phi)$, 
$\Delta S(\phi K^0_S, \eta' K^0_S) = \sin 2 \beta_{eff} - \sin 2 \beta$
and $S(K^{*0}\gamma) = \sin 2 \beta_{eff}$.}
\label{fig:comparison}
\end{figure}

We presented a selection of key measurements that could be performed
with an upgrade of the LHCb experiment, namely precision measurements
of the weak mixing phase \phis\ in tree decays and $b\to s$ transitions,
and of the CKM angle~$\gamma$. 
Many more measurements would also benefit
from a ten-fold increase in statistics. 
With 100~\fbinv\ of data, LHCb would be able measure 
$\Delta S(\phi K^0_S) = \sin 2 \beta_{eff} - \sin 2 \beta$ in \Bdtophiks\ with  
an estimated precision of 
$\sigma(\Delta S(\phi K^0_S)) \sim 0.04$. 
An estimated yield 
of 44k \Bdtokstarmumu\ events per year 
would allow probing right-handed currents
by analysing forward-backward asymmetries ($A_{FB}$) and transversity angles. 
In addition, the LHCb upgrade would allow the probeing of NP in charm decays
and to study lepton flavour violation in $\tau$ decays.

In Fig.~\ref{fig:comparison} we present a comparison of the expected sensitivities 
for the LHCb upgrade, scaled from 10 to 100~\fbinv, with 
sensitivities at a Super-B factory based on a $\rm 50\; ab^{-1}$ data sample~\cite{ref:superB}.
Both approaches have comparable sensitivities for many measurements such as 
$\gamma$ and $\sin 2\beta_{eff}$. LHCb will be limited in reconstructing
decays with (more than one) neutral final state particles (photon, $\pi^0$ or $\eta$). 
However only an upgrade of the LHCb experiment will allow an ultimate exploitation of  \Bs\ meson decays. 
While these estimates need to be confirmed with simulations at high
luminosities, 
an excellent science case can be written
for an LHCb upgrade.

\section{LHCb DETECTOR AND TRIGGER UPGRADE}
\label{sec:detector}

\subsection{Detector and trigger plans }

The trigger for the LHCb experiment comprises a first level trigger (L0) 
implemented in hardware and a Higher Level Trigger (HLT) running on a 
large CPU farm. The L0 trigger operates at 40 MHz and selects high transverse 
momentum objects in the electromagnetic and hadronic 
calorimeters ($e$, $\gamma$, charged hadrons) and muons in the muon system ($\mu$, $\mu \mu$).
Pile-up events are vetoed.
The output rate is limited to 1.1 MHz with $4\,\mu s$ latency.
Simulations show that the L0 muon trigger efficiency 
for reconstructable events
is around 90\% and that the 
output rate scales with luminosity up to $5\times 10^{32}\, \rm cm^{-2}\,s^{-1}$.
At the design luminosity, 
the L0 hadron trigger efficiencies are about 40\%, 
and the output rate increases only slightly with luminosity. 
The vertex detector (VELO) sensors~\cite{ref:lhcbreopt} 
undergo radiation damage and it is expected that 
these will need to be replaced when 6 to 8 \fbinv\ of luminosity has been collected.
This illustrates that the existing trigger and at least some of the sub-detector systems 
will not allow operating the LHCb experiment at ten times the design luminosity.

We have commenced studies which investigate how to upgrade the LHCb detector and triggers
such that the experiment can operate at luminosities ${\cal L} \sim 2\times 10^{33}\, \rm cm^{-2}\,s^{-1}$. 
Two scenarios are under consideration. A step-by-step approach would foresee to replace
the VELO sensors with improved radiation-hard detectors which will be read out 
at 40 MHz. This would allow the addition of a first-level detached vertex trigger at L0.
The Trigger Tracker 
in front of the LHCb dipole magnet could also be included. 
This trigger could be implemented in FPGAs, and it would need to be very fast and run within the 
L0 latency time. Other LHCb detector systems will also need to be upgraded when increasing the luminosity, 
either due to unacceptably large occupancies or radiation effects.
We are currently studying the luminosity limit of each LHCb sub-detector system.
Under discussion is the replacement of the central region of the RICH\,1 photon detectors,
an increase (decrease) of the Inner (Outer) Tracker area, and a replacement of the inner
region of the electromagnetic calorimeter. 

Another approach would be to change the read-out of all LHCb sub-detectors to 40~MHz. 
This has clear advantages as it would allow the implementation of
a L0 displaced vertex trigger in a CPU farm. In fact all trigger decisions would be 
software-based which allows flexibility.
However this approach requires a redesign of the front-end electronics 
which has implications for all sub-detector systems.
Besides the VELO, all silicon sensors of the Trigger and Inner Tracker and the RICH photon detectors 
would need to be replaced. 
The R\&D for the new front-end electronics will be able to profit from the developments 
of ATLAS and CMS for Super-LHC luminosities.

\subsection{Initial studies}

R\&D efforts have started on technologies for radiation-hard vertex detectors 
that will be able to operate in the radiation environments of the LHC and LHCb upgrade.
The detector sensors will need to be able to operate at radiation doses of 
about $10^{15}\; 1\; \rm MeV \; equivalent\; neutrons/cm^2$. 
Initial studies of Czochralski and $n$-on-$p$ sensors
irradiated up to $4.5 \times 10^{14}$ 24 GeV protons/cm$^2$ 
are promising and show that the charge collection efficiencies saturate at 
acceptable bias voltages~\cite{ref:parkes}.
Three-dimensional sensors are another alternative that could be investigated.

Two different vertex-detector geometries are envisaged. 
One is to shorten the strips, the other is to use pixels. 
Removing the RF foil that separates
the VELO sensors from the primary beam-pipe vacuum would reduce the radiation length before
the first measurement by 3\% and improve the proper time resolution of $B$ meson decays.

A preliminary study uses \Bstodsk\ decays simulated at 
a luminosity of $6\times 10^{32}\, \rm cm^{-2}\,s^{-1}$. 
Events with large numbers of interactions are employed to simulate 
larger effective luminosities up to $2\times 10^{33}\, \rm cm^{-2}\,s^{-1}$. 
In a first step, the existing HLT VELO trigger algorithm is added without modifications 
into the L0 trigger.
We find that the minimum bias rate keeps rising  with luminosity and saturates
the bandwidth well below our target luminosity. 
A better approach is to combine the L0 trigger with a detached vertex trigger 
and to read out at a 40\,MHz rate.
We require a transverse energy $E_T > 3\; \rm GeV$ from the hadron trigger and
combine this with a matched track that has a transverse momentum $p_T > 2$ GeV/$c$ and an 
impact parameter $\delta > 50 \mu \rm m$. 
In this combined trigger the minimum bias rate does not depend strongly on the luminosity and 
the triggered event yield scales linearly with the luminosity. In addition, the total trigger efficiency
is 60\% larger when compared with the existing baseline. 

\section{CONCLUSIONS}
\label{sec:conclusions}

We have presented a plan to upgrade the LHCb experiment after about five years of running. 
This programme is motivated by new physics beyond the Standard Model 
that can be studied with a ten times larger data sample. 
The LHCb upgrade would allow a precise measurement the weak mixing phase \phis\ 
at its Standard Model value and to probe new physics in $B$ meson decays at an 
unprecedented level of a few percent or better. 
Initial estimates show that an excellent science case can be developed.
The plan is to upgrade the LHCb detector such that it can operate at 
ten times the design luminosity, i.e. at ${\cal L} \sim 2\times 10^{33}\, \rm cm^{-2}\,s^{-1}$.
The upgraded experiment can run with or without an LHC luminosity upgrade.
The LHCb upgrade will  require a first-level detached vertex trigger. 
To achieve this, the vertex detector sensors will need to be replaced 
and many other sub-systems will need to be improved. 
Feasibility studies for physics  and trigger, and R\&D for 
detectors and front-end electronics are now required.


\end{document}